\documentclass[journal=jacsat,manuscript=article]{achemso}
\usepackage[version=3]{mhchem} 
\usepackage{siunitx}
\usepackage[dvipsnames]{xcolor}
\usepackage{amssymb}
\usepackage{pifont}
\usepackage{wrapfig}
\usepackage{hyperref}

\usepackage{amsmath}

\usepackage{tabularx,colortbl}
\usepackage{capt-of}
\usepackage{adjustbox}
\usepackage[font=normalsize,labelfont={bf},font={bf}]{caption}
\usepackage{graphicx}
\usepackage{subcaption}
\usepackage{tikz}
\usepackage{caption}
\usepackage{comment}
\usepackage{array}
\usepackage{multirow}
\usepackage[labelfont=md,textfont=md]{caption}
\usepackage{booktabs}
\usepackage{pgfplots}
\pgfplotsset{compat=1.18} 

\author{Radheesh Sharma Meda}
\affiliation[cheme]
{Department of Chemical Engineering, Carnegie Mellon University, 15213, USA}

\author{Amir Barati Farimani}
\email{barati@cmu.edu}
\affiliation[meche]
{Department of Mechanical Engineering, Carnegie Mellon University, 15213, USA}
\alsoaffiliation [biomed] {Department of Biomedical Engineering, Carnegie Mellon University, 15213, USA}
\alsoaffiliation [cheme] {Department of Chemical Engineering, Carnegie Mellon University, 15213, USA}
\alsoaffiliation [mld] {Machine Learning Department, Carnegie Mellon University, 15213, USA}

\title[An \textsf{achemso} demo]
{BAPULM: Binding Affinity Prediction using Language Models}

\begin{document}

 \begin{abstract}


\noindent 

Identifying drug-target interactions is essential for developing effective therapeutics. Binding affinity quantifies these interactions, and traditional approaches rely on computationally intensive 3D structural data. In contrast, language models can efficiently process sequential data, offering an alternative approach to molecular representation. In the current study, we introduce BAPULM, an innovative sequence-based framework that leverages the chemical latent representations of proteins via ProtT5-XL-U50 and ligands through MolFormer, eliminating reliance on complex 3D configurations. Our approach was validated extensively on benchmark datasets, achieving scoring power (R) values of 0.925 $\pm$ 0.043, 0.914 $\pm$ 0.004, and 0.8132 $\pm$ 0.001 on benchmark1k2101, Test2016\_290, and CSAR-HiQ\_36, respectively. These findings indicate the robustness and accuracy of BAPULM across diverse datasets and underscore the potential of sequence-based models \textit{in-silico} drug discovery, offering a scalable alternative to 3D-centric methods for screening potential ligands.

\end{abstract}

\section{Introduction} 
Developing novel therapeutics is essential for addressing extant diseases, newly emerging or untreated diseases, and future potential disorders that have yet to be identified\cite{Mollaei2024IDP-Bert:Models}. The recent COVID-19 pandemic has underscored the critical importance of rapid and innovative drug development to combat these unforeseen global challenges \cite{Blanchard2022LanguageInhibitors,Patil2023ForecastingModel}. In this pursuit, drugs, typically organic molecules composed of carbon-catenated structures (ligands), are stereoselectively designed to interact with specific amino acid motifs of their target proteins \cite{Mollaei2023UnveilingApproach,Du2016InsightsMethods}. These interactions are often mediated by non-covalent forces such as hydrogen bonds, van der Waals interactions, and electrostatic forces \cite{Adhav2024TheFunction}. Understanding the strength of these protein-ligand interactions, often represented by the equilibrium dissociation constant ($\text{K}_{\text{d}}$), is crucial to advance therapeutic development\cite{WangDeepDTAF:Affinity}. Spectroscopic techniques, including FTIR, NMR, UV-visible spectroscopy, and fluorescence, are employed to test potential ligands for specific proteins\cite{Kotting2013MonitoringSpectroscopy,Dalvit2023AffinityThem,Nienhaus2005ProbingSpectroscopy, Rossi2011AnalysisPolarization}. These methods capture conformational transitions within the secondary structure through vibrational bands, structural modifications through chemical changes, changes in absorbance due to the electronic environment, and alterations in fluorescence intensity upon protein-ligand binding, respectively\cite{Zhang2023HaPPy:Abstract,Qi2024MeasuringNMR}. 

In addition to these experimental approaches, computational methods such as molecular docking and molecular dynamics (MD) simulations have revolutionized affinity prediction by offering physical interpretability \cite{Zhao2020ExploringPrediction,Zheng2019OnionNet:Prediction}. While MD simulations accurately estimate binding affinities at the expense of higher compute power, molecular docking enables the exploration of large libraries of potential ligands, offering rapid virtual screening capabilities albeit reduced accuracy. Despite their limitations, these techniques laid the foundation for \textit{in silico} methods in drug discovery, paving the way for the adoption of deep learning models, which have achieved considerably higher predictive accuracy. 

Alongside molecular docking and simulations, 3D structure-based deep learning models adeptly capture the complex spatial features of protein-ligand interactions; however, they are inherently constrained by the dependence on high-resolution crystallographic data. In contrast, the emergence of large-scale datasets featuring sequential 1D representations of proteins and ligands enables the examination of the sequential molecular latent space for the screening of potential ligands\cite{Wang2022DLSSAffinity:Model,Zheng2019OnionNet:Prediction,Guntuboina2023PeptideBERT:Prediction}. With the availability of large-scale sequential datasets, researchers have developed advanced models such as transformers to leverage these data to produce more accurate affinity predictions. The transformer architecture inherently relies on the attention mechanism, which excels at comprehending sequential data. Language models leverage this architecture, using unsupervised pretraining to capture nuanced and comprehensive relationships within the data while encoding the sequences \cite{Elnaggar2021ProtTrans:Learning, Guntuboina2023PeptideBERT:Prediction,Kuan2024AbGPT:Modeling}. Elnaggar et al. pioneered the development of protein sequence-based language models such as ProtBERT, ProtAlbert, ProtElectra, and ProtT5, trained on expansive datasets UniRef, BFD comprising up to 393 billion amino acids. Interestingly, these models excel at attending to sequences that are spatially proximal, highlighting the importance of nearby amino acids over more distant ones \cite{VigBERTOLOGYMODELS}. Subsequently, ligand-specific encoder models such as ChemBerta and Molformer were engineered to encode the SMILES representation of organic molecules.

Building on these advancements, PLAPT successfully integrates BERT-based encoders for protein and ligand sequences to improve affinity predictions\cite{RosePLAPT:TRANSFORMERS}. However, the multimodal framework designed by Xu et al. demonstrates superior performance by incorporating additional binding pocket information through a residue graph network and employing cross-attention between the sequential and structural modalities. Yet, there remains an essential requirement for configurations that can achieve better predictive capabilities without the complications associated with the extensive data and computational demands of the MFE framework. The current study aims to address this research gap by exploring the synergistic utilization of pre-trained language models as a compelling alternative in the realm of protein-ligand binding affinity prediction. We present binding affinity prediction using language models (BAPULM), a framework that capitalizes on the integrated strengths of the ProtT5-XL-U50 \cite{Elnaggar2021ProtTrans:Learning} and Molformer \cite{Ross2021Large-ScaleProperties} encoder models to effectively estimate binding affinity with a predictive feedforward network. By utilizing these unsupervised pre-trained language models, BAPULM achieves high accuracy in binding affinity prediction while maintaining computational efficiency. BAPULM captures stereochemical molecular space and efficiently screens potential ligands, achieving state-of-the-art performance in predicting the binding affinity.

\section{Methods} 

\begin{figure}[ht]
    \centering
    \includegraphics[width=0.8\textwidth]{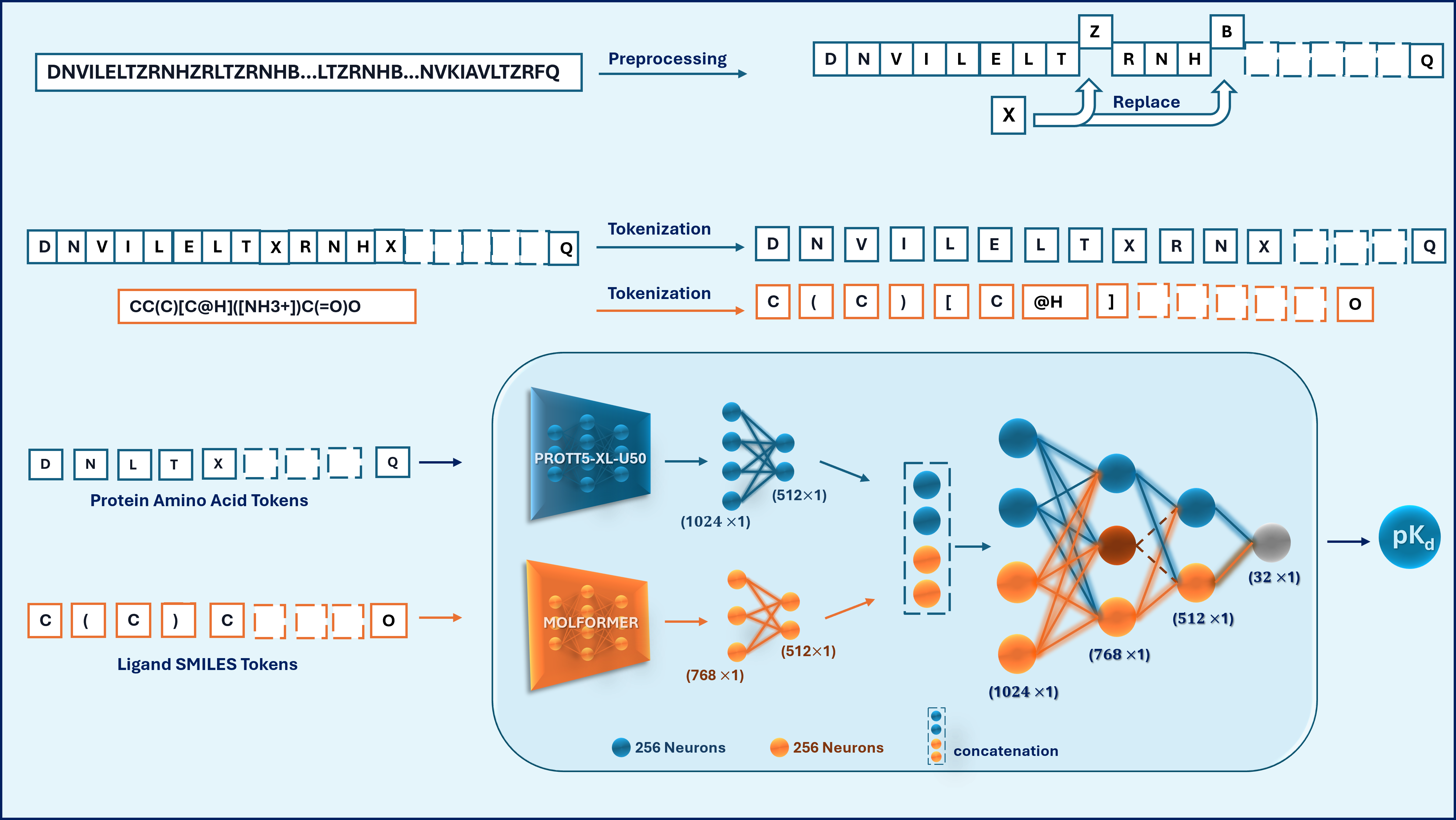}
    \caption{The overview of the BAPULM framework, which integrates the ProtT5-XL-U50 for protein sequnces and Molformer for ligand SMILES for feature extraction module while encoding the sequnces. These embeddings are aligned through projection layers and fed into a feed-forward predictive network to predict binding affinity.}
    \label{fig:framework}
\end{figure}

BAPULM was developed to utilize the functionality of encoder-based language models, which require simple 1D string expressions as input, such as protein amino acid sequences and ligand SMILES representation, to predict affinity as shown in Figure \ref{fig:framework}.

\subsection{Datasets}
\begin{figure}[ht]
    \centering
    \includegraphics[width=0.8\textwidth]{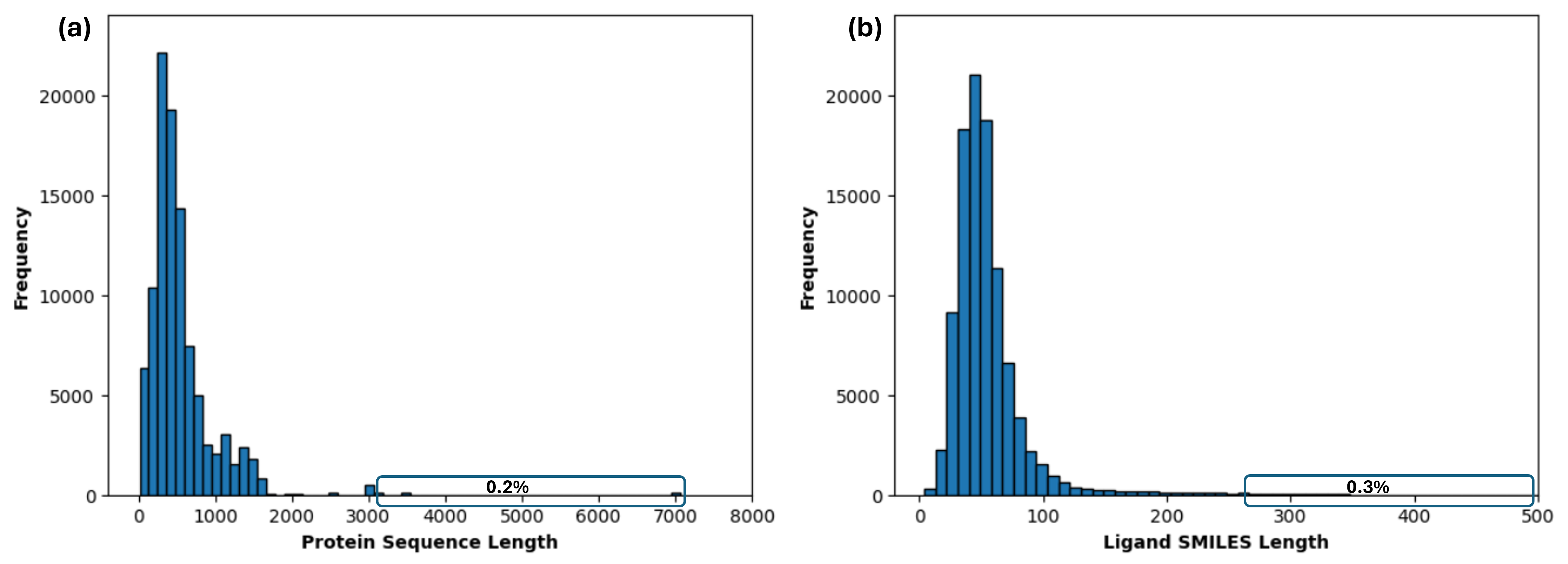}
    \caption{Distribution of (a) Protein sequence lengths range from 13 to 7073 amino acids, showing a skewed distribution with most sequences concentrated under 1000 amino acids. (b) Ligand SMILES string lengths range from 4 to 547 characters, also displaying a skewed distribution with the majority of strings being shorter than 100 characters.}
    \label{fig:distribution}
\end{figure}

The dataset employed to train BAPULM is the Binding Affinity Dataset \cite{Glaser2022BindingDataset} from the Hugging Face platform, which includes the curated pair of 1.9M unique set of protein-ligand complexes with the experimentally determined binding affinity $\text{pK}_{\text{d}}$. BAPULM operates on the subset of the first 100k aminoacid sequences, canonical smiles, and binding affinity ($\text{pK}_{\text{d}}$). Figure \ref{fig:distribution}. illustrates the distribution of (a) protein sequence length with only a tiny portion (~0.2\%) of the sequences with a length greater than 3200 and (b) ligand SMILES with a small fraction (~0.3\%) greater than 278.

A dataset of protein-ligand feature embeddings, $\text{pK}_{\text{d}}$, and normalized binding affinity was generated before model training using the encoder models described in Section 2.3. A split ratio of 90:10 was used to build training and validation sets, similar to the percentage employed in the previous work \cite{RosePLAPT:TRANSFORMERS}. Furthermore, the following benchmark datasets were acquired from the various works of literature: Benchmark1k2101 \cite{RosePLAPT:TRANSFORMERS}, Test2016\_290 \cite{Jin2023CAPLA:Mechanism}, and CSAR-HiQ\_36 \cite{Dunbar2011CSARComplexes} to evaluate BAPULM.  Every benchmark dataset was meticulously examined to ensure no overlapping with the training dataset \cite{RosePLAPT:TRANSFORMERS}. 

\subsection{PreProcessing}

Macromolecules built from the same set of 20 amino acid repeating units to form unique sequences are proteins. As a part of preprocessing, the protein sequences were separated by spaces into single characters (A-Z) describing the monomeric residuals and to standardize the input sequences, the non-essential amino acids Asparagine (B), Selenocysteine (U), Glutamic acid (Z), and Pyrrolysine (O) were replaced by employing the substitution code 'X' \cite{RosePLAPT:TRANSFORMERS,Elnaggar2021ProtTrans:Learning}.  The canonical SMILES captures the structural stereochemistry of the organic micro/macro molecules, ensuring a unique expression for every individual molecule, enabling a standardized representation.

\subsection{Model Architecture}
BAPULM's architecture consists of two robust components that are synergistically utilized to predict $\text{pK}_{\text{d}}$. Primarily, the feature encoding module harnessed the potency of ProtT5-XL-U50 for protein sequence and Molformer for ligand SMILES to generate consolidated vectors in latent space that constitute all the characteristic information about the proteins and ligands known as feature embeddings, which were subsequently utilized in the forthcoming module.

\subsubsection{Protein-ligand feature embedding}
The BAPULM model integrates the ProtT5-XL-U50 model, which is founded on the T5 model \cite{Raffel2020ExploringTransformer}, and differentiates itself from BERT by employing a unified transformer architecture (both encoder and decoder) while capturing the biophysical features of amino acids and the language of life \cite{Elnaggar2021ProtTrans:Learning,Raffel2020ExploringTransformer}. The preprocessed sequences are transformed into tokens following a comprehensive tokenization procedure, as mentioned in ProtTrans \cite{Elnaggar2021ProtTrans:Learning}.  This method involves padding and truncating the sequence to a maximum length of 3200, also a norm followed by previous work \cite{RosePLAPT:TRANSFORMERS}, generating a list of token IDs and their attending attention mask.  Subsequently, the tokens were passed to the encoder, and a mean pooling operation was performed on the last layer to generate fixed 1024-dimensional feature embeddings, enabling a comprehensive understanding of the protein sequences with variable lengths.  BAPULM further leverages Molformer, a state-of-the-art transformer-based encoder model, which effectively captures the spatial connection between the atoms in the SMILES sequence \cite{Ross2021Large-ScaleProperties}.  The canonical SMILES of ligands were tokenized while processed through padding and truncating to an utmost length of 278, including micro and macromolecule ligands.  The mean pooler output from the encoder was a 768-dimensional embedding vector containing the stereochemical features of the ligand molecule. A detailed breakdown of the lengths of the protein-ligand sequences is available in Supporting Information Table \ref{tab:protein_sequences} and \ref{tab:ligand_molecules}.

Therefore, the protein sequence was encoded into a 1024-dimensional embedding space while the ligand smiles to a 768-dimensional vector. To hereafter utilize these in the prediction module, both sets of feature vectors were then separately projected onto a lower-dimensional (512) latent space through a linear transformation employing ReLU (rectified linear unit) activation. These consolidated 512-dimensional feature vectors were concatenated to form a 1024-dimensional input vector to the feed-forward network.

\subsubsection{Feed-Forward Predictive Network}
The concatenated 1024-dimensional combined feature vector was passed through four ReLU-activated linear layers, as shown in Figure \ref{fig:framework}. Before passing through the linear layers, the mini-batches of combined feature embeddings underwent batch normalization to improve training stability by reducing the internal covariance shift\cite{Ioffe2015BatchShift}. Dropout was also applied to avert overfitting and create a robust model. The last layer output of the model yielded a normalized scalar value of the binding affinity($\text{pK}_{\text{d}}$).

\subsection{Training and Evaluation Metrics}
The previously generated feature dataset was utilized to train BAPULM, employing Mean Squared Error(MSE) as a loss function, which estimates the average squared difference between the actual $\text{pK}_{\text{d}}$ and predicted affinity as shown below:

\begin{equation}\label{eq1}
\text{MSE} = \frac{1}{n} \sum_{i=1}^{n} \left( \text{pK}_{\text{d,true}, i} - \text{pK}_{\text{d,pred}, i} \right)^2
\end{equation}

This loss function was optimized utilizing the Adam optimizer to update the model's weights. The training process was executed on an Nvidia RTX 2080 Ti with 11GB of memory and completed in approximately four minutes. Additionally, the training hyperparameters are provided in the Supporting Information Table \ref{tab:hyperparameters}. 

To estimate the efficacy of BAPULM in predicting the negative log of the binding affinity dissociation constant ($\text{pK}_{\text{d}}$) between protein-ligand complexes, we used the following evaluation metrics: Mean Absolute Error (MAE), Root Mean Squared Error (RMSE), and Pearson correlation coefficient (R) as shown in the equations \ref{eq2}, \ref{eq3}, \ref{eq4}, where $\text{pK}_{\text{d}_{ true}}$, $\text{pK}_{\text{d}_{ pred}}$ correspond to the experimental and predicted affinities.

\begin{equation}
\label{eq2}
\text{MAE} = \frac{1}{n} \sum_{i=1}^{n} \left| \text{pK}_{\text{d,true}, i} - \text{pK}_{\text{d,pred}, i} \right|
\end{equation}

\begin{equation}
\label{eq3}
    \text{RMSE} = \sqrt{\frac{1}{n} \sum_{i=1}^{n} \left( \text{pK}_{\text{d,true}, i} - \text{pK}_{\text{d,pred}, i} \right)^2}
\end{equation}

\begin{equation}
\label{eq4}
    R = \frac{\sum_{i=1}^{n} \left( \text{pK}_{\text{d,true}, i} - \mu_{\text{pK}_{\text{d,true}}} \right) \left( \text{pK}_{\text{d,pred}, i} - \mu_{\text{pK}_{\text{d,pred}}} \right)}{\sqrt{\sum_{i=1}^{n} \left( \text{pK}_{\text{d,true}, i} - \mu_{\text{pK}_{\text{d,true}}} \right)^2 \sum_{i=1}^{n} \left( \text{pK}_{\text{d,pred}, i} - \mu_{\text{pK}_{\text{d,pred}}} \right)^2}}
\end{equation}

These metrics are widely adopted in regression studies and were established in published literature \cite{Zheng2019OnionNet:Prediction,Jin2023CAPLA:Mechanism,Zhang2023HaPPy:Abstract,Xu2024Surface-basedPrediction}. In particular, the person correlation coefficient (R) was considered as one of the scoring power metrics in evaluating the performance \cite{Zheng2019OnionNet:Prediction}. Again, both RMSE and MAE were employed to provide a comprehensive understanding of performance, as RMSE is optimal for errors with a normal distribution. In contrast, MAE is better suited for errors with a Laplacian distribution \cite{Hodson2022Root-mean-squareNot}.  Since these metrics evaluate predicted and experimental $\text{pK}_{\text{d}}$ values, the model's output was denormalized onto the same scale as the experimental affinity to assess the performance.

\section{Results and Discussion}

\begin{figure}[ht]
    \centering
    \includegraphics[width=\textwidth]{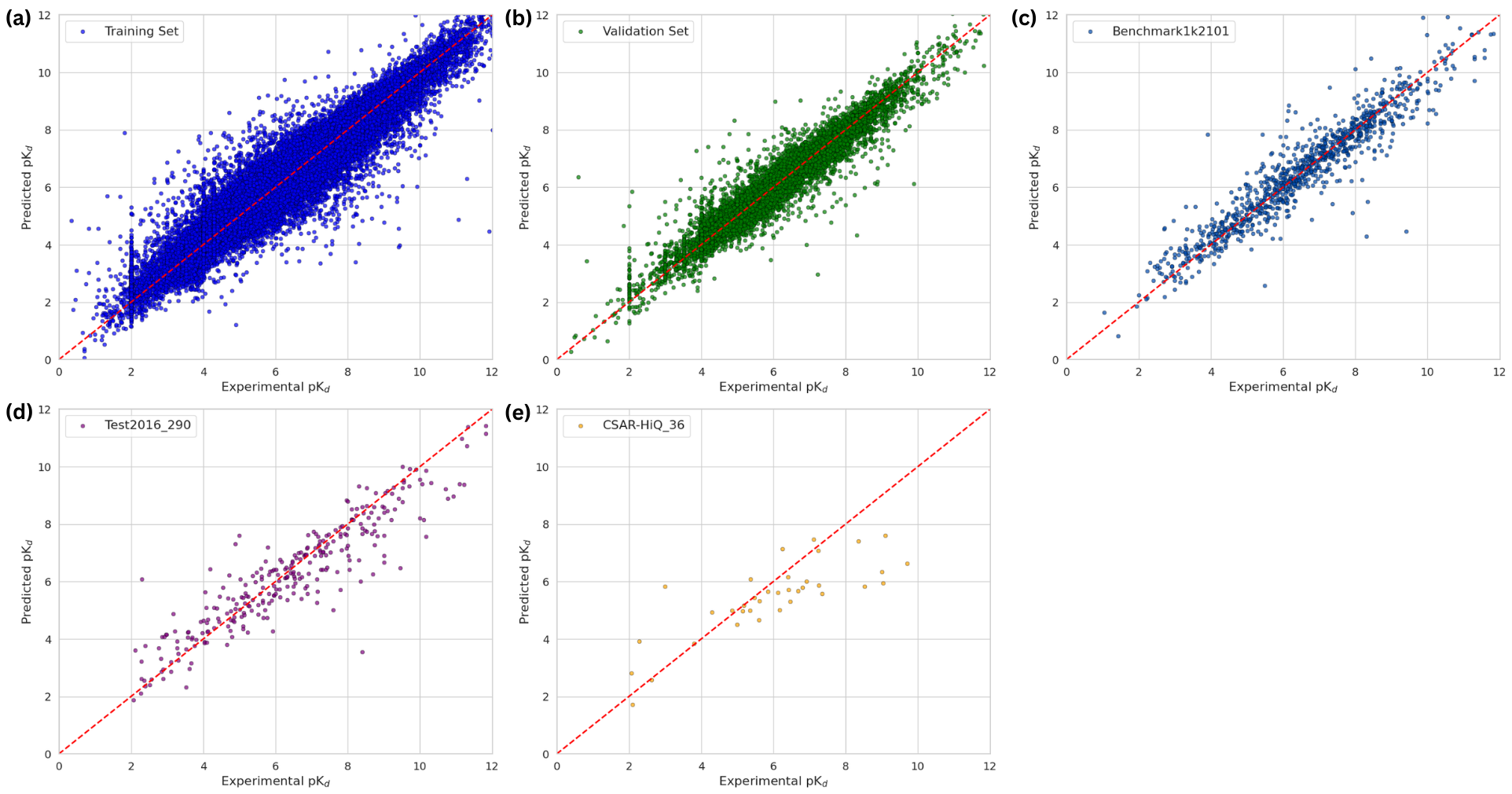}
    \caption{ Evaluation of BAPULM on multiple datasets where the scatter plots depict the correlation between predicted and experimental $\text{pK}_{\text{d}}$ values. The datasets represented include the (a) Training ,(b) Validation (c) Benchmark1k2101,(d) Test2016\_290, and (e)CSAR\-HiQ\_36. }
    \label{fig:parityplots}
\end{figure}

BAPULM's unique ability to predict binding affinity originates from the inherent nature of its architecture, which effectively captures the intricate features of protein sequences and ligand molecular structures. As shown in Table \ref{tab:train}, BAPULM constantly displayed an improvement in each metric compared to PLAPT\cite{RosePLAPT:TRANSFORMERS}, demonstrating its exceptional performance. Notably, BAPULM achieved a higher person correlation coefficient (R) with an increase of 9.6\% (0.970) and 40.7\% (0.960) on training and validation datasets, respectively, indicating a robust correlation between predicted and experimental $\text{pK}_{\text{d}}$ values. Also, the consolidated clustering of points along the identity line in the parity plots, as displayed in Figure \ref{fig:parityplots}(a,b), corroborates with the higher correlation coefficient. 

\begin{table}[ht!]
\centering
\caption{Evaluation Metrics for BAPULM and PLAPT on Training and Validation Datasets}
\begin{tabular}{lccccc}
\toprule
\textbf{Dataset} & \textbf{Model} & \textbf{R \(\uparrow\)} & \textbf{MSE \(\downarrow\)} & \textbf{RMSE \(\downarrow\)} & \textbf{MAE \(\downarrow\)} \\
\midrule
\multirow{2}{*}{Train} & BAPULM (this study) & 0.970 & 0.157 & 0.397 & 0.245 \\
& PLAPT & 0.886 & 0.586 & 0.765 & 0.756 \\
\midrule
\multirow{2}{*}{Validation} & BAPULM (this study) & 0.960 & 0.177 & 0.421 & 0.248 \\
& PLAPT & 0.683 & 1.466 & 1.211 & 0.949 \\
\bottomrule
\end{tabular}
\label{tab:train}
\end{table}

Furthermore, BAPULM exhibited remarkably lower error metrics, with a drop of 73.2\%, 48.1\%, and 67.6\% in MSE (0.157), RMSE (0.397), and MAE (0.245), respectively, on the training data  Similarly, on the validation data, the model showed a decline of 87.9\% in MSE (0.177), 65.3\% in RMSE (0.421), and 73.9\% in MAE (0.248), underscoring its predictive capability. This significant improvement across both training and validation datasets demonstrated the ability of the model to comprehensively capture the underlying interactions between the proteins and ligands, facilitating accurate predictions. 

Moreover, BAPULM's predictive ability was further validated on three distinct benchmark datasets, where it was compared to current state-of-the-art models, as shown in Table \ref{tab:benchmark}. The evaluation metrics in Table \ref{tab:benchmark}  are computed as the mean and standard deviation, estimated using different seed values (2102, 256, 42), to accurately reflect the model's performance during inference on test datasets with the trained model weights. Accordingly, on the benchmark2k1k dataset, BAPULM demonstrates improved evaluated values compared to PLAPT, with an increase in the R-value of 4.76\% and a drop in RMSE, MAE by 19.1\% and 37.2\%. 

\begin{table}[ht!]
\centering
\caption{Model Performance on Various Benchmark Datasets}
\resizebox{\textwidth}{!}{ 
\begin{tabular}{lccccccc}
\toprule
\textbf{Dataset} & \textbf{Model} & \textbf{Data Representation} & \textbf{Feature extraction} & \textbf{R \(\uparrow\)} & \textbf{RMSE \(\downarrow\)} & \textbf{MAE \(\downarrow\)} \\
\midrule
\multirow{2}{*}{benchmark1k2101} & \textbf{BAPULM} (this study) & seq + smiles\_canonical & ProtT5-XL-U50 + Molformer & \textbf{0.925} $\pm$ \textbf{0.043} & \textbf{0.745} $\pm$ \textbf{0.236} & \textbf{0.432} $\pm$ \textbf{0.013} \\
& PLAPT & seq + smiles\_canonical & ProtBert + ChemBerta & 0.883 & 0.922 & 0.688 \\
\midrule
\multirow{7}{*}{Test2016\_290} & \textbf{BAPULM} (this study) & seq + smiles\_canonical & ProtT5-XL-U50 + Molformer & \textbf{0.914} $\pm$ \textbf{0.004} & \textbf{0.898} $\pm$ \textbf{0.0172} & \textbf{0.645} $\pm$ \textbf{0.0166} \\
& MFE  & Protein seq, 3D structure + ligand graph & Multimodal between seq, structure + ligand graph &  0.851 & 1.151 & 0.882 \\
& PLAPT & seq + smiles\_canonical & ProtBert + ChemBerta & 0.845 & 1.196 & 0.906 \\
& CAPLA & protein seq, ligand smiles + binding pocket & 1D convolution block + Cross attention (pocket/ligand) & 0.843 & 1.200 & 0.966 \\
& DeepDTAF & Protein seq, ligand smiles + binding pocket & 1D Conv, 1D Conv + 3 Conv layers for binding pocket & 0.789 & 1.355 & 1.073 \\
& OnionNet & Protein-ligand 3D grid & 3D Conv + Neural Attention &  0.816 & 1.278 & 0.984 \\
\midrule
\multirow{5}{*}{CSAR-HiQ\_36} & \textbf{BAPULM} (this study) & seq + smiles\_canonical & ProtT5-XL-U50 + Molformer & \textbf{0.8132} $\pm$ \textbf{0.012} & \textbf{1.328} $\pm$ \textbf{0.020} & \textbf{1.029} $\pm$ \textbf{0.022} \\
& affinity\_pred & - & - & 0.774 & 1.484 & 1.176 \\
& PLAPT & seq + smiles\_canonical & ProtBert + ChemBerta & 0.731 & 1.349 & 1.157 \\
& CAPLA & seq, smiles + binding pocket & 1D convolution block + Cross attention (pocket/ligand) & 0.704 & 1.454 & 1.160 \\
& DeepDTAF & seq + smiles & 1D CNN on seq and smiles & 0.543 & 2.765 & 2.318 \\
\bottomrule
\end{tabular}
}
\label{tab:benchmark}
\end{table}

Xu et al.\cite{Xu2024Surface-basedPrediction} developed a multimodal feature extraction (MFE) framework that employed the following feature extraction module involving 1D protein sequence, binding pocket surface through point cloud, 3D structural features, and the ligand molecular graph. It slightly outperformed PLAPT on the Test\_2016 dataset by 0.6\% improvement in correlation coefficient (R) while reducing the RMSE and MAE by 3.8\% and 2.6\%, becoming the current state-of-the-art affinity prediction model. However, BAPULM leveraging ProtT5-XL-U50, Molformer substantially outperformed MFE's performance by 7.4\%, 21.8\%,26.7\% in R (0.914) , RMSE(0.898) and MAE (0.642), respectively. Additionally, BAPULM surpassed both sequence and structure-based models on every metric. It outperformed CAPLA\cite{Jin2023CAPLA:Mechanism} by 8.4\% in R, 25.2\% in RMSE, and 32.2\% in MAE. Against DeepDTAF\cite{WangDeepDTAF:Affinity}, BAPULM showed a higher linear correlation value with an increase of 15.9\%, reduced RMSE by 33.7\%, and decreased MAE by 39.9\%. Furthermore, compared to OnionNet\cite{Zheng2019OnionNet:Prediction}, it achieved a 12\% higher R-value, a lower RMSE, and an MAE of 29.7\% and 34.5\%, respectively. This implies that BAPULM was successfully able to capture the linear relationship between $\text{pK}_{\text{d}}$ (experimental) and $\text{pK}_{\text{d}}$ (predicted), alongside being more accurate by achieving lower RMSE and MAE values.

Finally, on the CSAR-HiQ\_36 dataset, BAPULM yet again proved its exceptional predictive ability. Unlike PLAPT, BAPULM was able to capture the identity relationship between predicted and actual binding affinity, besides being accurate \cite{RosePLAPT:TRANSFORMERS}. BAPULM achieved a notable scoring power value of 0.813, denoting an 11.2\% improvement over PLAPT and 5.1 \%  against affinity\_pred\cite{Blanchard2022LanguageInhibitors}. Similarly, the percentage improvement on the other two metrics was greater (MAE: 12.5\%, RMSE: 10.5\%) than PLAPT's advancement over affinity\_pred (MAE: 1.62\%, RMSE:9.10\%). Additionally, BAPULM outperforms other sequence-based models on R, RMSE, and MAE against CAPLA by 15.25\%,8.67\%,11.29\%, and over DeepDTAF by 48.7\%, 51.96\%, 55.59\%, respectively.

\begin{figure}[ht]
    \centering
    \includegraphics[width=\textwidth]{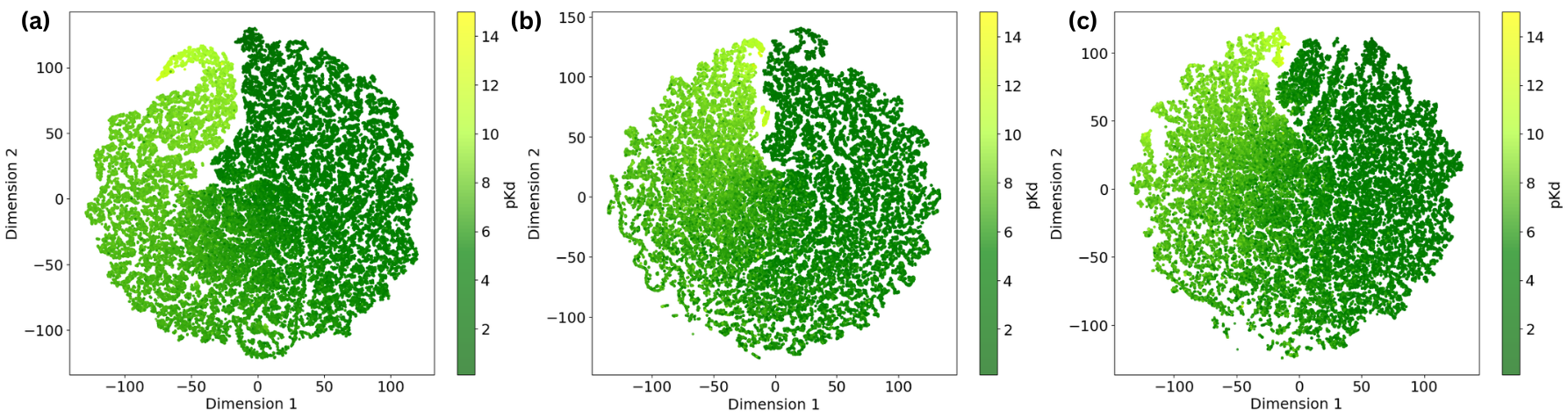}
    \caption{Embedding visualizations of protein-ligand binding affinity mapped onto features extracted from (a) BAPULM, (b) ProtBert \& Molformer, and (c) ProtBert \& ChemBerta, illustrating the latent space representations of each configuration on train dataset.}
    \label{fig:tSNE}
\end{figure}

Furthermore, to gain insights into BAPULM’s excellent correlation capabilities, features from the penultimate layer were extracted and utilized to generate t-distributed Stochastic Neighbor Embedding (t-SNE) visualizations. t-SNE is a statistical method that maps high-dimensional data to a lower-dimensional space, conserving the local structure and enabling the visualization in a lower dimension\cite{Badrinarayanan2024Multi-Peptide:Properties}. To understand the influence of encoder-based language models in predicting binding affinity, we employed the combination of transformer encoders, such as protBERT, ChemBERTa, and Molformer, within the same model architecture, assessing their ability to capture the binding affinity between protein-ligand complexes effectively. BAPULM demonstrates a clear and distinct gradient transition in the t-SNE visualization, indicating a strong correlation between the latent representations of protein-ligand complexes and their binding affinities. In contrast, the distribution for the ProtBERT and MolFormer models is more dispersed, with less noticeable separation of embeddings based on $\text{pK}_{\text{d}}$ values. Similarly, the t-SNE visualization for ProtBERT and ChemBERTa shows a partial gradient transition but with some overlap between high-affinity and low-affinity complexes. Although both ProtBERT \& MolFormer and ProtBERT \& ChemBERTa exhibit some clustering of complexes according to $\text{pK}_{\text{d}}$, the clustering is much more prominent in BAPULM.  This is attributed to using rotary positional embeddings in Molformer during pretraining, enabling it to learn spatial relationships within the ligand. The synergistic combination of Molformer with ProtT5-XL-U50 in BAPULM effectively captured the binding affinity correlation, resulting in a clear and distinct separation of protein-ligand complexes in the t-SNE visualization. This separation is characterized by a smooth color gradient, indicating BAPULM's ability to distinguish between complexes with varying binding affinities.

\section{Conclusion}
This study introduces a sequence-based machine-learning model, BAPULM, that leverages transformer-based language models ProtT5-XL-U-50 and Molformer to predict protein-ligand binding affinity. BAPULM effectively captures the latent features of protein-ligand complexes without relying on structural data, enabling a robust representation by harnessing the inherent information in biochemical sequences. This approach significantly enhances predictive accuracy while reducing computational complexity. The integration of Molformer with rotary positional encoding enhanced BAPULM's ability to comprehend the stereochemistry of ligands without requiring detailed  3D configurations to demonstrate superior performance across diverse benchmarks. Our t-SNE visualizations reveal that synergistic integration of these encoders displayed a distinct clustering of complexes according to binding affinity, substantiating BAPULM's predictive capability. This framework presents an efficient alternative to conventional structure-based models, demonstrating the potential of using sequence-based models for rapid virtual screening.


\section{Data and Software Availability}

The necessary code and data used in this study can be accessed here:
\url{https://github.com/radh55sh/BAPULM.git}

\begin{acknowledgement}

We acknowledge the contributions of various individuals and organizations that have made this study possible. This includes the providers of the datasets used in our research, the developers of PyTorch, and the teams behind ProtT5-XL-U50 and Molformer. 

\end{acknowledgement}

\bibliography{reference}
\newpage
\section{Supporting Information}
\subsection*{Sequence Distributions}

Table \ref{tab:protein_sequences}, \ref{tab:ligand_molecules} present the detailed length distributions of protein sequences and ligand molecules in our dataset.

\begin{table}[h!]
    \centering
    \begin{tabular}{lcr}
        \toprule
        \textbf{Length Range} & \textbf{Number of Protein Sequences} \\
        \midrule
        1--1000   & 88,485 \\
        1001--2000 & 10,598 \\
        2001--3200 & 706 \\
        3201--4000 & 123 \\
        4001--7073 & 88 \\
        \bottomrule
    \end{tabular}
    \caption{Distribution of Protein Sequences by Length Range}
    \label{tab:protein_sequences}
\end{table}

\begin{table}[h!]
    \centering
    \begin{tabular}{lcr}
        \toprule
        \textbf{Length Range} & \textbf{Number of Ligand Molecules} \\
        \midrule
        1--100   & 94,831 \\
        101--200 & 4,085 \\
        201--278 & 753 \\
        279--478 & 330 \\
        479--547 & 1 \\
        \bottomrule
    \end{tabular}
    \caption{Distribution of Ligand Molecules by Length Range}
    \label{tab:ligand_molecules}
\end{table}

\newpage
\subsection*{Hyperparameters}

Table \ref{tab:hyperparameters} summarizes the key hyperparameters, detailing essential configurations utilized for training the model.

\begin{table}[h!]
    \begin{tabular}{lcr}
        \toprule
        \textbf{Hyperparameters} & \textbf{Value} \\
        \midrule
        Seed & 2102 \\
        Loss Function & MSE \\
        Optimizer & Adam \\
        Learning Rate & 1e-3 \\
        Batch size & 256 \\
        Epochs & 60 \\
        Scheduler & ReduceLROnPlateau \\
        Scheduler Patience & 5 \\
        Scheduler Factor & 0.2 \\
        \bottomrule
    \end{tabular}
    \caption{BAPULM model hyperparameters}
    \label{tab:hyperparameters}
\end{table}

\end{document}


\subsection*{Sequence Distributions}

Table \ref{tab:protein_sequences}, \ref{tab:ligand_molecules} present the detailed length distributions of protein sequences and ligand molecules in our dataset. The majority of proteins (88,485) fall within the 1-1000 length range, while most ligands (94,831) have lengths between 1-100, providing comprehensive coverage of biologically relevant molecular sizes.

\begin{table}[h!]
    \centering
    \begin{tabular}{lcr}
        \toprule
        \textbf{Length Range} & \textbf{Number of Protein Sequences} \\
        \midrule
        1--1000   & 88,485 \\
        1001--2000 & 10,598 \\
        2001--3200 & 706 \\
        3201--4000 & 123 \\
        4001--7073 & 88 \\
        \bottomrule
    \end{tabular}
    \caption{Distribution of Protein Sequences by Length Range}
    \label{tab:protein_sequences}
\end{table}

\begin{table}[h!]
    \centering
    \begin{tabular}{lcr}
        \toprule
        \textbf{Length Range} & \textbf{Number of Ligand Molecules} \\
        \midrule
        1--100   & 94,831 \\
        101--200 & 4,085 \\
        201--278 & 753 \\
        279--478 & 330 \\
        479--547 & 1 \\
        \bottomrule
    \end{tabular}
    \caption{Distribution of Ligand Molecules by Length Range}
    \label{tab:ligand_molecules}
\end{table}

\subsection*{Hyperparameters}

Table \ref{tab:hyperparameters} summarizes the key hyperparameters used for training the BAPULM, detailing essential configurations for achieving optimal performance.

\begin{table}[h!]
    \centering
    \begin{tabular}{lcr}
        \toprule
        \textbf{Hyperparameters} & \textbf{Value} \\
        \midrule
        Seed & 2102 \\
        Loss Function & MSE \\
        Optimizer & Adam \\
        Learning Rate & 1e-3 \\
        Batch size & 256 \\
        Epochs & 60 \\
        Scheduler & ReduceLROnPlateau \\
        Scheduler Patience & 5 \\
        Scheduler Factor & 0.2 \\
        \bottomrule
    \end{tabular}
    \caption{Summary of hyperparameters for training the BAPULM model.}
    \label{tab:hyperparameters}
\end{table}